\documentclass[final,3p,times,twocolumn]{elsarticle}

\usepackage{amssymb}
\usepackage{amsmath}

\usepackage{lineno}
\usepackage{graphicx}
\usepackage{here}
\usepackage{float}
\usepackage[legacycolonsymbols]{mathtools}
\usepackage[english]{babel}
\usepackage{xcolor}
\usepackage{bm}
\usepackage{siunitx}

\usepackage{color}
\usepackage{ulem}
\usepackage{soul}
\usepackage{ifthen}

\newcounter{verbosity}
\newcommand\temp{\bgroup\markoverwith{\textcolor{red}{\rule[.5ex]{2pt}{0.4pt}}}\ULon}
\newcommand{\Add}[1]{\textcolor{black}{#1}}       
\newcommand{\Addrev}[1]{\textcolor{black}{#1}}  
\ifthenelse{\value{verbosity} > 1}
{
\newcommand{\Erase}[1]{\if0{#1}\fi}	     
\newcommand\EraseL[1]{\if0{#1}\fi}          
}{
\newcommand{\Erase}[1]{\if0{#1}\fi}	            
\newcommand\EraseL[1]{\if0{#1}\fi}              
}

\journal{Acta Astronautica}

\begin{document}

\begin{frontmatter}



\title{Autonomous optical navigation for {$\text{DESTINY}^\text{+}$}: Enhancing misalignment robustness in flyby observations with a rotating telescope}   


\author[label1]{Takayuki Hosonuma\corref{cor1}}
\cortext[cor1]{Corresponding author}
\ead{hosonuma@space.t.u-tokyo.ac.jp}
\author[label2]{Takeshi Miyabara}
\author[label2]{Naoya Ozaki}
\author[label3]{Ko Ishibashi}
\author[label2]{Yuta Suzaki}
\author[label3]{Peng Hong}
\author[label2]{Masayuki Ohta}
\author[label2]{Takeshi Takashima}

\affiliation[label1]{organization={The University of Tokyo},
            addressline={Bunkyo-ku},
            city={Tokyo},
            postcode={113-8656},
            country={Japan}}
\affiliation[label2]{organization={Institute of Space and Aeronautical Science (ISAS), Japan Aerospace Exploration Agency (JAXA)},
            addressline={Sagamihara},
            city={Kanagawa},
            postcode={252-5210},
            country={Japan}}
\affiliation[label3]{organization={Planetary Exploration Research Center},
            addressline={Chiba Institute of Technology, Narashino},
            city={Kanagawa},
            postcode={275-0016},
            country={Japan}}

\begin{abstract}
{$\text{DESTINY}^\text{+}$} is an upcoming JAXA Epsilon medium-class mission to flyby multiple asteroids including Phaethon. 
As an asteroid flyby observation instrument, a telescope mechanically capable of single-axis rotation, named TCAP, is mounted on the spacecraft to track and observe the target asteroids during flyby.
As in past flyby missions utilizing rotating telescopes, TCAP is also used as a navigation camera for autonomous optical navigation during the closest-approach phase.
To mitigate the degradation of the navigation accuracy, past missions performed calibration of the navigation camera's alignment before starting optical navigation.
However, such calibration requires capturing large number of images by the navigation camera and downlinking them to the ground station.
In the case of small spacecraft with limited link capacity, such as {$\text{DESTINY}^\text{+}$}, such ground-in-the-loop calibration requires significant operational time to complete and imposes constraints on the operation sequence.
From the above background, the {$\text{DESTINY}^\text{+}$} team has studied the possibility of reducing operational costs by allowing TCAP alignment errors to remain.
This paper describes an autonomous optical navigation algorithm robust to the misalignment of rotating telescopes, proposed in this context.
In the proposed method, the misalignment of the telescope is estimated simultaneously with the spacecraft’s orbit relative to the flyby target. 
To deal with the nonlinearity between the misalignment and the observation value, the proposed method utilizes the unscented Kalman filter, instead of the extended Kalman filter widely used in past studies.
The proposed method was evaluated with numerical simulations on a PC and with hardware-in-the-loop simulation, taking the Phaethon flyby in the {$\text{DESTINY}^\text{+}$} mission as an example.
In the example case, the misalignment-induced navigation accuracy degradation of \EraseL{4.6 km}\Add{\SI{4.6}{\kilo \metre}}-\(1\sigma\) can be reduced to \Erase{0.1 km}\Add{\SI{0.1}{\kilo \metre}}-\(1\sigma\) by the proposed method.
The required time to run one-cycle of the navigation process on the onboard computer for the {$\text{DESTINY}^\text{+}$} mission is less than \Erase{0.18 s}\Add{\SI{0.18}{\second}}.
These results validate that the proposed method can mitigate the misalignment-induced degradation of the optical navigation accuracy with reasonable computational costs suited for onboard computers.
\end{abstract}




\end{frontmatter}


\section{Introduction} 

Flyby observation is one of the useful approaches to explore small bodies such as comets and asteroids. 
In flyby observation, spacecraft capture high-resolution images of target small bodies at closest approach. 
This is advantageous considering the low fuel consumption compared to the other approaches such as rendezvous and touch down, which enables exploring small bodies using small/nano spacecraft \cite{Funase2016,Machuca2019,Turan2022,Pugliatti2023} with limited fuel budget. 
Recently, several plans for flyby observations of multiple small bodies have been proposed \cite{Ozaki2022-1, Levison2021, Ozaki2022-2, Chabot2024}. 
Thus, flyby observation is expected to play an important role as a cost-effective method for exploring small bodies in the future.

There are two technical challenges in accomplishing precise flyby observation that arise from the fast motion of the target body relative to the spacecraft. 
The first problem is related to the attitude maneuverability of the spacecraft. 
To capture images of the target body during flyby, the spacecraft is required to align its line of sight (LoS) to the target body, which moves with a high angular velocity relative to the spacecraft. 
Typically, the required angular velocity varies from several deg/s to several tens of deg/s, which is difficult to achieve only with spacecraft attitude maneuvers. 
The second problem is related to the estimation accuracy of the spacecraft position relative to the target body. 
Even if the spacecraft has high maneuverability, it is difficult to precisely align its LoS to the target body without knowing the relative direction of the target. 
Since the target's relative direction rapidly changes during flyby, the spacecraft should autonomously update its knowledge of the target’s direction.

\begin{figure}[!tp]
    \centering
    \includegraphics[scale=0.55]{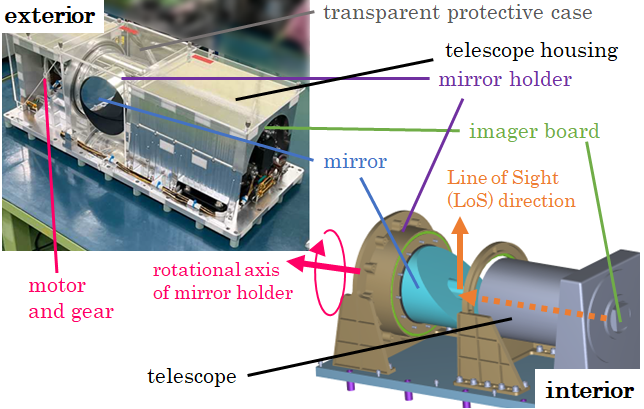}
    \caption{Example of rotating telescope (exterior view and interior CAD image of TCAP for $\text{DESTINY}^\text{+}$ mission)}
    \label{fig:rottel}
\end{figure}

\begin{figure}[!tp]
    \centering
    \includegraphics[scale=0.8]{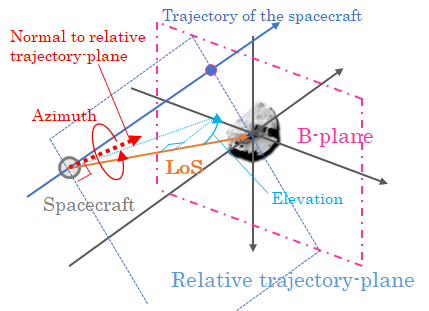}
    \caption{Example of flyby geometry utilizing a rotating telescope.}
    \label{fig:geometry}
\end{figure}

To address the first issue, about dealing with the limitation on spacecraft attitude maneuverability, rotating telescopes integrated with drive mechanisms have been used in several previous and ongoing flyby missions \cite{Funase2016, Toyota2023, Newburn}.
Rotating telescopes comprise a telescope, a bending mirror within a mirror holder attached to the front of the aperture of the telescope, and drive mechanics (motor and gear) to rotate the mirror holder. 
By rotating the mirror holder, the LoS of the telescope rotates around the rotational axis of the mirror holder.
Fig. \ref{fig:rottel} depicts an example of the rotating telescope, called the Telescopic Camera for Phaethon (TCAP) \cite{Hong}, which is under development for the Demonstration and Experiment of Space Technology for Interplanetary Voyage with Phaethon Flyby and Dust Science ($\text{DESTINY}^\text{+}$) mission \cite{Toyota2023, Ozaki2022-1}.
$\text{DESTINY}^\text{+}$ is a \Erase{480 kg}\Add{\SI{480}{\kilo \gram}} spacecraft being developed in line with JAXA's vision to realize low-cost and high-frequency deep space exploration utilizing small deep space probes \cite{Kawakatsu}.
For its nominal science mission, the $\text{DESTINY}^\text{+}$ spacecraft will perform a flyby observation of the active asteroid Phaethon utilizing TCAP. The spacecraft will fly by the asteroid at a closest distance of \Erase{500 km}\Add{\SI{500}{\kilo \metre}} with a relative velocity of \Erase{33 km/s}\Add{\SI{33}{km/s}} in the nominal case. 
Thus, the spacecraft is required to change its LoS direction with the maximum angular rate of \Erase{3.8 deg/s}\Add{\SI{3.8}{deg/s}} to track and observe the asteroid during the flyby.
Since the required angular rate exceeds the maneuverability of the spacecraft, the spacecraft drives the mirror holder of TCAP with single-axis motor, instead of performing the fast attitude maneuver.
To do so, as described in Fig. \ref{fig:geometry}, the spacecraft adjusts the elevation angle with its Attitude and Orbit Control System (AOCS) to align the rotational axis of TCAP normal to the relative trajectory plane. If the rotational axis of TCAP coincides with the vertical direction of the relative trajectory plane, the relative motion remains only in the rotational plane of TCAP. Therefore, by adjusting the azimuth angle via the single-axis rotational motion of TCAP, the target asteroid can be tracked without performing the fast attitude maneuver.
Hence, the rotating telescope can moderate the requirements for the attitude maneuverability of the spacecraft.

To address the second issue, about the estimation of the relative motion, past missions have generally adopted onboard optical navigation, in which the observation value is the centroid position of the target body on images obtained by navigation cameras \cite{Bhaskaran, Okada, Ariu, Kaluthantrige2023, Bellerose2024}.
Hence, several studies have been performed to improve optical navigation accuracy during flyby observation. 
For example, Bhaskaran \cite{Bhaskaran} described an optical navigation algorithm in which the attitude determination error and optical distortions were considered. Okada \cite{Okada} showed that shading of the target body can cause observation errors in optical navigation and described the mitigation of this effect. 
Ariu \cite{Ariu} suggested the non-uniform property of observation noise covariance during the flyby and proposed an algorithm considering the non-uniform property. In past missions utilizing the rotating telescopes, the relative trajectory of the spacecraft was estimated with onboard optical navigation using the rotating telescopes as navigation cameras \cite{Funase2016, Newburn, Lessac2024}. 
Similary, TCAP will be used as a navigation camera in the $\text{DESTINY}^\text{+}$ mission.
TCAP will be utilized for two different types of optical navigation: ground-in-the-loop navigation and onboard autonomous navigation.
The former starts several days before the closest approach and lasts until several hours before the closest approach, while the latter is performed a few hours before and after the closest approach.
The reason for using these two different navigation methods stems from the limited link capacity of the spacecraft. 
During several days before and after the closest approach, the data downlink rate from the spacecraft is expected to be a few kbps, while the TCAP image data size is at least several Mbyte per image. 
It is therefore expected that it will take several hours to downlink an image to the ground station.
Against this background, it is essential to perform the onboard navigation several hours before the closest approach.

However, using the rotating telescope as a navigation camera can cause further accuracy degradation due to misalignment. 
As described earlier, the rotating telescope comprises several mechanical parts that tend to be misaligned to the telescope. 
If the LoS direction of the telescope is displaced from the expected direction, it can cause unexpected observation errors in optical navigation, as described in Section 4.2 of this paper.
Nevertheless, few studies have been reported from the viewpoint of mitigating the degradation of optical navigation accuracy due to misalignments. 
One of the reasons for this is, by performing alignment calibration operations before starting the optical navigation, such accuracy degradation can be mitigated.
However, such calibration requires capturing large number of images as reference data and downlinking them to the ground station \cite{Lessac2024, Rush2022}.
In the case of small spacecraft missions with limited link capacity, such as the {$\text{DESTINY}^\text{+}$} mission, such ground-in-the-loop calibration requires significant operational time to complete and imposes constraints on the sequence of operation. 
Performing the calibration during the transfer phase from the Earth to the destination is an option to avoid this problem as it ensures a longer operational time to perform the calibration. 
However, the misalignment may change during the transfer phase as the thermal distortion on the spacecraft will change depending on solar distance and spacecraft attitude relative to the Sun.
Therefore, it is preferable to perform the calibration as close as possible to the closest approach timing.
On the other hand, such operation is difficult for the {$\text{DESTINY}^\text{+}$} mission due to the limited link capacity of a few kbps during the closest approach phase.
Thus, autonomous calibration of the misalignment during the closest approach is desired.

To this end, this paper proposes an autonomous optical navigation algorithm intended to apply several hours before and after the closest approach that can mitigate the accuracy degradation due to telescope misalignment.
In the proposed method, the misalignment of the telescope is estimated simultaneously with the spacecraft’s position relative to the flyby target during closest approach. 
A similar algorithm is already applied to the Lucy mission \cite{good}, which performs flyby observation with a telescope mounded on a two-axis gimbal.
This algorithm models the telescope misalignment as a pair of two bias angles (\(\alpha\), \(\beta\)) and estimates them simultaneously with the spacecraft’s position relative to the flyby target during closest approach. 
However, for flyby observation with a single-axis rotating telescope, such as the $\text{DESTINY}^\text{+}$ mission, this approach is not sufficient considering the purpose of the navigation.
As described earlier with Fig. \ref{fig:geometry}, the important purpose of navigation in the $\text{DESTINY}^\text{+}$ mission is to provide information for precise target tracking by aligning the TCAP rotational plane to the trajectory plane with attitude control.
However, as discussed later in Section 2, the rotation plane with misalignments is not flat but conical, and the LoS error in normal to the trajectory plane varies with the TCAP rotational angle, which is not constant bias.
Since this LoS error changes rapidly as the telescope rotates at high speeds, estimating the error as an instantaneous bias error is not sufficient to keep up with changes in the error.
In order to track this LoS error, it is effective to specify parameters to express the error as a function of the telescope rotational angle and to make them the estimation values, thereby reducing the time variation of the estimation values.
Therefore, in the proposed method, the telescope misalignment is decomposed in more detail and is represented by seven parameters to describe the LoS error in a predictable manner as a function of the telescope rotational angle. 
These seven parameters are estimated simultaneously with the spacecraft’s position relative to the flyby target during closest approach, utilizing unscented Kalman filter.

The remainder of this paper is organized as follows. 
First, the misalignment model of the rotating telescope is described in Section 2 based on geometrical insights. 
The relationship between the misalignment and the LoS error is also discussed in detail.
Then, the developed optical navigation algorithm considering the misalignment is introduced in Section 3. 
The evaluation of the effectiveness of the developed algorithm is presented in Section 4 with Monte-Carlo simulations and comparisons with a conventional optical navigation algorithm not considering the misalignment. 
Finally, in Section 5, an implementation of the developed algorithm on an onboard computer (OBC) with hardware-in-the-loop simulation (HILS) is presented to highlight its practicality as flight software.

\section{Misalignment Model of Rotating Telescope} 

This section derives the mathematical model of the misaligned rotating telescope. 
First, seven parameters representing the misalignment are identified with geometrical insight. 
Utilizing these parameters, the mathematical model to describe the LoS direction of the misaligned rotating telescope is obtained.

\subsection{Parameterization of misalignment} 

As described in Section 1, the rotating telescope comprises three sub-components: telescope, bending mirror in mirror holder, and motor. 
The misalignment of the rotating telescope comprises the displacement between these sub-components. 
In addition, the displacement between the telescope and the spacecraft body also occurs. 
Therefore, the misalignment of the rotating telescope comprises the following four components (A) to (D). 
To describe these misalignments, the base coordinate of the telescope is introduced in the following. In this coordinate, the X- and Y-axes are the horizontal and vertical directions of the imager respectively, and the Z-axis is aligned to the optical axis, which is normal to the imager plane.
In the following, Fig. \ref{fig:def-a-misalign} describes the misalignment (A) and Figs. \ref{fig:def-b-misalign} to \ref{fig:def-d-misalign} describe the misalignment (B) to (D), with mainly focusing on the relative attachment of the mirror and its holder described in Fig. \ref{fig:rottel} with respect to the imager of the telescope.

\begin{enumerate}
    \item[(A)] \textbf{\textit{Misalignment between the base coordinate of the telescope and spacecraft body}}
    
    ~~~~This type of misalignment is intended for mounting errors of the telescope with respect to the spacecraft body. 
    Since the alignment of the telescope with respect to the spacecraft can be defined as the Euler angles between the coordinate fixed to the telescope and the body coordinate of the spacecraft, type-A misalignment can be parameterized with three small Euler angles, \(\phi_A\), \(\theta_A\), and \(\psi_A\), as described in Fig. \ref{fig:def-a-misalign}.
    In this figure, \(X_{b}\), \(Y_{b}\) and \(Z_{b}\) represent the base vectors of the spacecraft body coordinate.
    \begin{figure}[!t]
        \centering
    \includegraphics[scale=0.62]{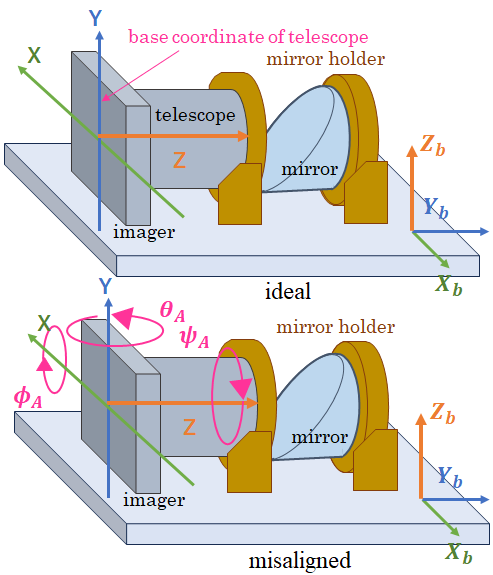}
    \caption{Definitions of Type-A misalignment}
    \label{fig:def-a-misalign}
    \end{figure}
    
    \item[(B)] \textbf{\textit{Misalignment of the rotational axis of the mirror}}

    ~~~~As described in Fig. \ref{fig:def-b-misalign}, the rotational axis of the mirror holder is expected to coincide with the optical axis of the telescope in the ideal condition. 
    Type-B misalignment represents the tilt angle of the rotational axis of the mirror holder with respect to the optical axis of the telescope. 
    Thus, type-B misalignment can be parameterized with two angles: the tilt angle of the rotational axis \(\delta_B\) and the tilt direction angle \(\phi_B\).
   \begin{figure}[!tpb]
    \centering
      \includegraphics[scale=0.7]{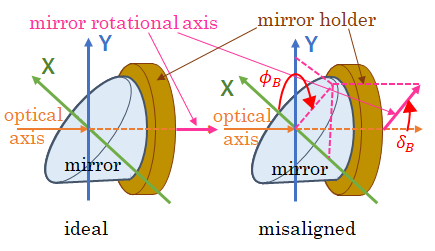}
    \caption{Definitions of Type-B misalignment}
    \label{fig:def-b-misalign}
    \end{figure}

    \item[(C)] \textbf{\textit{Misalignment of the tilt angle of the mirror}}
    
    ~~~~As described in Fig. \ref{fig:def-c-misalign}, in the ideal condition, the mirror is expected to be attached to the mirror holder with a tilt angle of \Erase{45 degrees}\Add{\SI{45}{deg}} with respect to the optical axis. 
    Type-C misalignment represents the attachment error angle of the mirror with respect to the mirror holder, which is hereafter denoted as \(\delta_C\).
    \begin{figure}[!pbt]
	\centering
	  \includegraphics[scale=0.7]{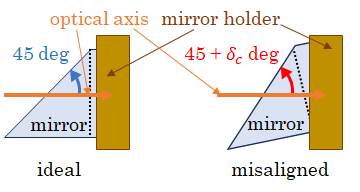}
	\caption{Definitions of Type-C misalignment}
	\label{fig:def-c-misalign}
    \end{figure}

    \item[(D)] \textbf{\textit{Offset of the rotational angle of the mirror}}
    
    ~~~~As described in Fig. \ref{fig:def-d-misalign}, in the ideal condition, the normal vector of the bending mirror lies in the \Erase{X}\Add{Y}Z plane and its projection vector on the XY plane points +Y direction when the rotational angle of TCAP is zero. 
    Type-D misalignment represents the offset of this projection vector from +Y direction when the rotational angle of TCAP is zero. In short, Type-D misalignment represents the offset of the rotational angle of the mirror from zero.
    Thus, type-D misalignment can be represented with one angle \(\delta_D\).
    \begin{figure}[!pbt]
	\centering
	  \includegraphics[scale=0.7]{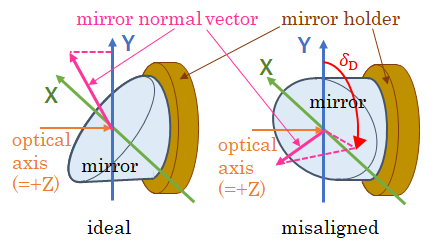}
	\caption{Definitions of Type-D misalignment}
	\label{fig:def-d-misalign}
    \end{figure}
    
\end{enumerate}

\begin{table}[!b]
\caption{\Add{Definition of misalignment parameters}}\label{tbl_misalign}
\centering
\begin{tabular}{p{1.5cm} p{5cm}}
\hline
\Erase{p}\Add{P}arameter & \Erase{d}\Add{D}escription \\ 
\hline
\(\phi_{A}\), \(\theta_{A}\), \(\psi_{A}\) & misalignment of telescope with respect to spacecraft body \\
\(\delta_{B}\), \(\phi_B\) & tilt angle of rotational axis and tilt direction \\
\(\delta_{C}\) & error angle of mirror relative to mirror holder \\
\(\delta_{D}\) & offset of rotation angle of mirror from zero \\
\hline
\end{tabular}
\end{table}

In addition to the misalignments (A)--(D), there can also be translational offsets of each element, but while these affect the imaging performance of the telescope, they do not affect the directional error of the LoS.
Hence, this study adopts the seven parameters described above to represent the misalignment of the rotating telescope.

\subsection{LoS direction without misalignment} 

To develop the mathematical model of the misalignment effect, this section first derives the mathematical model of the LoS direction “without” considering the misalignment effect.

\begin{figure}
	\centering
	\includegraphics[scale=0.6]{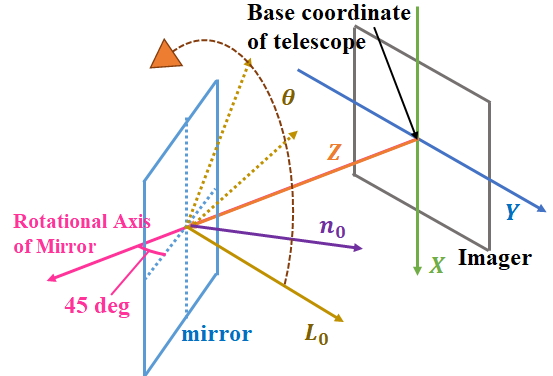}
	\caption{LoS model "without" considering misalignment}
	\label{fig:Losnominal}
\end{figure}

The LoS after mirror reflection can be expressed mathematically by focusing on the normal vector of the mirror plane and LoS before mirror reflection \cite{Golub}. 
From the definition of the base coordinate of the telescope, the mirror normal vector in the base coordinate \(\textbf{n}_0\) is
\begin{equation}
    \label{eq:n0nominal}
    \textbf{n}_0 = [0, \cos(\frac{\pi}{4}), -\sin(\frac{\pi}{4})]^{T}
\end{equation}
Additionally, from the definition of the base coordinate, the rotational axis of the mirror \(\epsilon\) is
\begin{equation}
    \label{eq:enominal}
    \boldsymbol{\epsilon} =[\epsilon_{x}, \epsilon_{y}, \epsilon_{z}]^{T} = [0, 0, 1]^{T}
\end{equation}
When the mirror holder rotates, the normal vector of the mirror plane \(\textbf{n}_0\) rotates around \(\epsilon\) by the rotational angle of the mirror holder \(\theta\).
Thus, the normal vector of the mirror plane after the rotation of the mirror holder becomes \(\textbf{n}_{(\theta)}\), which is a function of \(\theta\) described as follows:
\begin{equation}
    \label{eq:nnominal}
    \textbf{n}_{(\theta)} =  \Tilde{q}_{(\theta)} * \textbf{n}_0 * q_{(\theta)}
\end{equation}
where,
\begin{equation}
    \label{eq:qrot}
    q_{(\theta)} = [-\epsilon_{x}\sin{\frac{\theta}{2}}, -\epsilon_{y}\sin{\frac{\theta}{2}}, -\epsilon_{z}\sin{\frac{\theta}{2}}, \cos{\frac{\theta}{2}}]^{T}
\end{equation}
and “*” denotes the Hamilton product of the vectors. \(\Tilde{q}_{(\theta)}\) represents the conjugation of \(q_{(\theta)}\).
From the definition of the base coordinate, the LoS before the mirror reflection \(\textbf{L}_0\) is
\begin{equation}
    \textbf{L}_0 = [0, 0, 1]^{T}
\end{equation}
Therefore, from the Householder model of mirror reflection 
 \cite{Golub} , the LoS after mirror reflection \(\textbf{L}\) is
\begin{eqnarray}
    \label{eq:Lnominal}
    \textbf{L} &=& \textbf{L}_0 - 2(\textbf{L}_0\cdot\textbf{n}_{(\theta)})\textbf{n}_{(\theta)}\\
    \label{eq:Lnominal2}
    &=& [-\sin{\theta}, \cos{\theta}, 0]^{T}
\end{eqnarray}
Thus, in the ideal condition, the LoS of the telescope scans in the vertical plane with respect to the rotation axis, as described in Fig. \ref{fig:Losnominal}.

\subsection{LoS direction with misalignment} 
In this section, the LoS model of the telescope described in Eq. (\ref{eq:Lnominal}) is modified to incorporate the misalignment effect.
From the type-C misalignment, the mirror normal vector at the origin of the rotation angle \(\textbf{n}_0\) becomes
\begin{eqnarray}
    \label{eq:n-missed}
    \textbf{n}_0 &=& [0, \cos{(\frac{\pi}{4}+\delta_c)}, -\sin{(\frac{\pi}{4}+\delta_c)}]\\
    &\approx&
    [0, \frac{1-\delta_c}{\sqrt{2}}, -\frac{1+\delta_c}{\sqrt{2}}]^{T}
\end{eqnarray}
Moreover, from the type-B\Erase{-} misalignment, the rotational axis of the mirror \(\epsilon\) becomes
\begin{eqnarray}
    \label{eq:e-missed}
    \epsilon &=& [\sin{\delta_B}\cos{\phi_B}, \sin{\delta_B}\sin{\phi_B}, \cos{\delta_B}]^{T}\\
    &\approx&
    [\delta_B\cos{\phi_B}, \delta_B\sin{\phi_B}, 1]^{T}
\end{eqnarray}
From the type-D\Erase{-} misalignment, the rotation angle of the mirror holder \(\theta\) is replaced with \(\eta\), which is defined as follows:
\begin{equation}
    \label{eq:theta-missed}
    \theta \rightarrow \theta + \delta_D \equiv \eta
\end{equation}
By substituting Eqs. (\ref{eq:n-missed}) \(\sim\) (\ref{eq:theta-missed}) into Eqs. (\ref{eq:nnominal}), (\ref{eq:qrot}), and (\ref{eq:Lnominal}), the LoS direction with the misalignment effect is obtained as follows
\begin{equation}
\label{eq:L-missed}
\begin{split}
    &\textbf{L} = \textbf{L}_0 - 2(\textbf{L}_0\cdot\textbf{n}_{(\eta)})\textbf{n}_{(\eta)}\\
    &\approx
\begin{bmatrix}
    -\sin{\eta} - \delta_B\sin{\frac{\eta}{2}}\left(\sin{\left(\phi_B-\frac{\eta}{2}\right)} -\sin{\eta}\cos{\left(\phi_B - \frac{\eta}{2}\right)} \right) \\
    \cos{\eta} + \delta_B\sin{\frac{\eta}{2}}\left(\sin{\left(\phi_B+\frac{\eta}{2}\right)} +\sin{\eta}\cos{\left(\phi_B + \frac{\eta}{2}\right)}\right) \\ 
    -2\delta_C-2\delta_B\sin{\frac{\eta}{2}}\cos{\left(\phi_B - \frac{\eta}{2} \right)}
\end{bmatrix}
\end{split}
\end{equation}
Comparing Eq. (\ref{eq:L-missed}) to the ideal LoS direction expressed in Eq. (\ref{eq:Lnominal2}) suggests that the misalignment causes the directional error of the LoS in both the in-rotational- and out-rotational-plane directions, and the error varies with the rotational angle of the mirror.

\begin{figure}[!tp]
	\centering
	\includegraphics[scale=0.65]{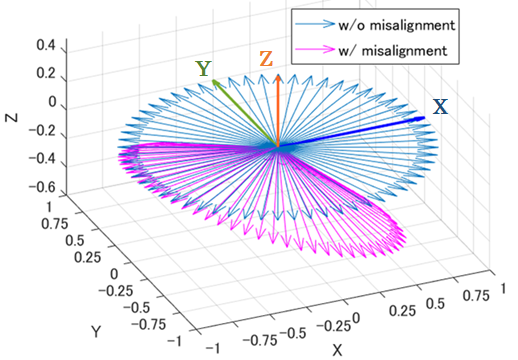}
	\caption{Rotational motion of LoS direction}
	\label{fig:LoS3D}
\end{figure}

\begin{figure}[!tp]
	\centering
	\includegraphics[scale=0.6]{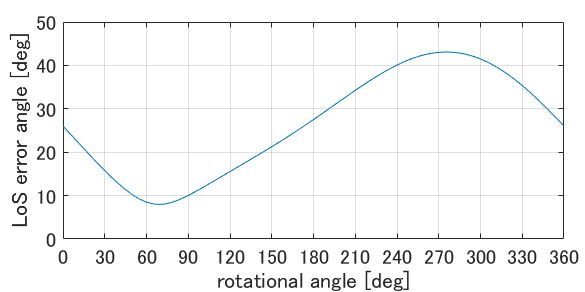}
	\caption{LoS error angle due to misalignment}
	\label{fig:LoSError}
\end{figure}

To illustrate the misalignments' effect, the trajectory of the LoS while the mirror is rotating one revolution is calculated for the cases with and without the misalignments. 
To make the effect easier to see, the amount of the misalignments are set extremely large here. 
Specifically, \(\delta_{b}\), \(\phi_{b}\) and \(\delta_{c}\) are set to \Erase{10 deg}\Add{\SI{10}{deg}}, while the other misalignments are set to \Erase{0 deg}\Add{\SI{0}{deg}}. 
Fig. \ref{fig:LoS3D} describes the trajectories of LoS in the base coordinate of the telescope.
In the absence of misalignments, the trajectory remains in the XY-plane as described with blue vectors, whereas in the presence of misalignments, the trajectory deviates from the XY-plane and moves in the conical plane as described with magenta vectors.
Moreover, the amounts of deviation from the XY-plane depend on the rotation angle of the mirror, as described in Fig. \ref{fig:LoSError}.
Thus, as it noted in introduction of this paper, it is difficult to describe the rotational plane of LoS in terms of only two constant bias angles.

\section{Optical Navigation Algorithm}
This section describes the optical navigation algorithm considering the effect of the misalignment of the rotating telescope. 
First, the state equation and observation equation are derived leveraging the misalignment model described in the previous section. Then, the filter equations for the optical navigation are described.

\subsection{Orbit and state dynamics model in navigation filter} 
In accordance with previous studies \cite{Bhaskaran, Okada, Ariu, Bhaskaran1998}, in this study, the relative motion of the spacecraft with respect to the target body was approximated as linear uniform motion in the STR coordinate \cite{Jah2002, Vallado}. 
The detailed definition of STR coordinate is described in \ref{app1}.
\begin{table}[!b]
\caption{Orbital elements used in approximation error analysis}\label{tbl_oe}
\centering
\begin{tabular}{p{2.8cm} p{3.8cm}}
\hline
Elements & \Erase{v}\Add{V}alue \\ 
\hline
Semi major axis & \Erase{1.020953 au}\Add{\SI{1.020953}{au}} (spacecraft) \\
& \Erase{1.271435 au}\Add{\SI{1.271435}{au}} (asteroid) \\
Eccentricity & 0.120506 (spacecraft) \\
& 0.889728 (asteroid) \\
Inclination & \Erase{24.04032 deg}\Add{\SI{24.04032}{deg}} (spacecraft) \\
& \Erase{30.49029 deg}\Add{\SI{30.49029}{deg}} (asteroid) \\
Right Ascension of & \Erase{1.633242 deg}\Add{\SI{1.633242}{deg}} (spacecraft) \\
Ascending Node & \Erase{311.7874 deg}\Add{\SI{311.7874}{deg}} (asteroid) \\
Argument of & \Erase{133.5874 deg}\Add{\SI{133.5874}{deg}} (spacecraft) \\
Periapsis & \Erase{270.9515 deg}\Add{\SI{270.9515}{deg}} (asteroid) \\
Mean Anomaly & \Erase{318.5986 deg}\Add{\SI{318.5986}{deg}} (spacecraft) \\
& \Erase{335.7213 deg}\Add{\SI{335.7213}{deg}} (asteroid) \\
\hline
\end{tabular}
\end{table}

\begin{figure}[!b]
	\centering
	\includegraphics[scale=0.65]{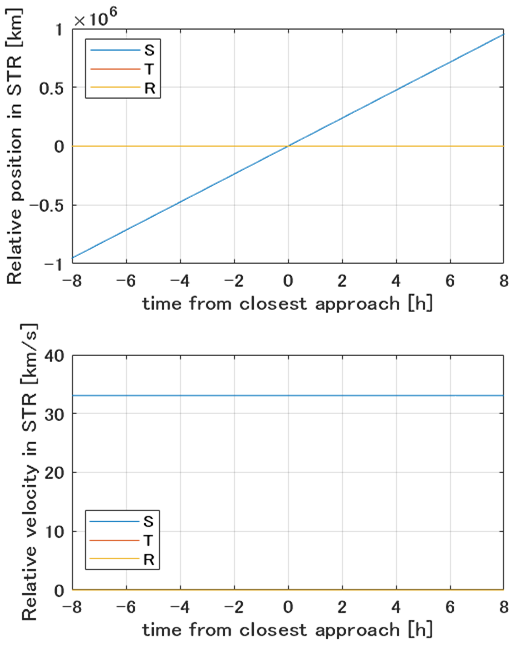}
	\caption{history of spacecraft's relative position and velocity in STR coordinate (upper:position, lower:velocity)}
	\label{fig:PosVel_STR}
\end{figure}
\begin{figure*}[!t]
	\centering
	\includegraphics[scale=0.67]{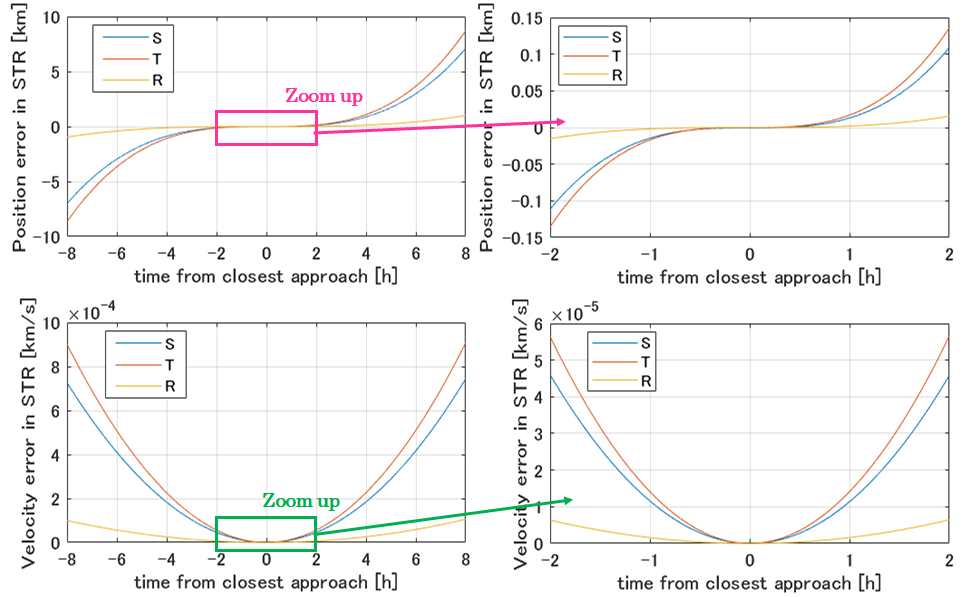}
	\caption{history of spacecraft's position and velocity error by constant velocity linear motion approximation (upper:position error, lower:velocity error)}
	\label{fig:Liner_ApproxDyn_Err}
\end{figure*}
To evaluate the error in the above approximation, an error analysis is performed on the Phaethon flyby case in the {$\text{DESTINY}^\text{+}$} mission.
In this analysis, first, the position and velocity of the spacecraft and asteroid are calculated as the two-body problem in the heliocentric inertial frame.
The gravitational perturbation from the asteroid is ignored, since the mass of the asteroid is negligibly small. 
Then, from the calculated results, the spacecraft's position and velocity relative to the asteroid are computed and converted to that in STR coordinate.
The orbital elements of the spacecraft and the asteroid used in this analysis at \Erase{8 h}\Add{\SI{8}{h}} before the closest approach are summarized in Table \ref{tbl_oe}.
Fig. \ref{fig:PosVel_STR} describes the history of the spacecraft's position and velocity relative to the asteroid in STR coordinate, during several hours before and after the closest approach timing.
The obtained spacecraft's relative trajectory in STR coordinate is compared to the approximated linear uniform motion to evaluate the approximation error.
The linear uniform motion is calculated by propagating the spacecraft's relative position forward and backward in time with the constant relative velocity from the closest approach timing.
Fig. \ref{fig:Liner_ApproxDyn_Err} describes the history of the approximation error on the spacecraft's relative position and velocity in STR coordinate.
As described in this figure, the approximation error in the relative position decreases to less than \Erase{10 km}\Add{\SI{10}{km}} at \Erase{8 h}\Add{\SI{8}{h}} before the closest approach, less than \Erase{0.13 km}\Add{\SI{0.13}{km}} at \Erase{2 h}\Add{\SI{2}{h}} before, and less than \Erase{0.02 km}\Add{\SI{0.02}{km}} at \Erase{1 h}\Add{\SI{1}{h}} before, which is negligibly small compar\Addrev{e} to the uncertainties of about \Erase{100 km}\Add{\SI{100}{km}} in the initial orbit determination prior to the autonomous navigation.    
From these results, it is confirmed that the spacecraft trajectory in STR coordinate can be approximated as a linear uniform motion for a few hours before and after the closest approach timing, which is assumed to be the main application period of the proposed method.
This approximation is advantageous for implementing the onboard software, as it can efficiently reduce computational costs. 

Thus, the spacecraft position \(\textbf{r}_{(t)}\) with respect to the target body at an arbitrary time \(t\) during several hours before and after the closest approach timing can be approximated as Eq. (\ref{eq:lin-mortion}).
\begin{equation}
    \label{eq:lin-mortion}
     \textbf{r}_{(t)} = \textbf{r}_{(t_0)} + \textbf{v}_{(t_0)}(t-t_0) + \Delta \textbf{r}_{(t)}
\end{equation}
where, \(t_0\) is the base time for the optical navigation, \(\textbf{r}_{(t_0)}\) is the relative position of the spacecraft at the base time, and \(\textbf{v}_{(t_0)}\) is the relative constant velocity of the spacecraft with respect to the target. 
This study assumes that \(\textbf{r}_{(t_0)}\) and \(\textbf{v}_{(t_0)}\) are obtained from delta differenced one-way ranging (DDOR) navigation and ground-in-the-loop optical navigation, which are conducted prior to optical navigation. 
Hence, the proposed method aims to estimate the position error \(\Delta\textbf{r}_{(t)}\), which is not eliminated in the prior navigational operations and evolve\Add{s} in time with the velocity error, while minimizing the degradation of the estimation accuracy induced by the misalignment of the telescope.

To achieve this, this study adopts a \Erase{10}\Add{ten}-dimensional state vector \(\textbf{x}\) comprising the current spacecraft relative position \(\textbf{r}_{(t)}\) and seven misalignment parameters described in the previous section, as shown in Eq. (\ref{eq:state_vec}).
\begin{equation}
\label{eq:state_vec}
    \textbf{x} = [\textbf{r}^{T}_{(t)}, \phi_A, \theta_A, \psi_A, \delta_B, \phi_B, \delta_C, \delta_D]^{T}
\end{equation}
In Eq. (\ref{eq:state_vec}), the spacecraft's relative velocity is not explicitly included in the state vector.
In the case of the {$\text{DESTINY}^\text{+}$} mission, the spacecraft's velocity is estimated by DDOR with an accuracy of \Erase{1 m/s}\Add{\SI{1}{m/s}}-\(3\sigma\), prior to the autonomous navigation.
Considering the flight time of a few hours from the final DDOR timing to the closest approach timing, the relative position error due to this velocity error is less than several km, which is much smaller than the residual position error of about \Erase{100 km}\Add{\SI{100}{km}}. Therefore, similar to the previous studies \cite{Bhaskaran, good, Bhaskaran1998}, the relative velocity is not included in the state vector. However, the position error partially derives from this velocity error is taken into account as a process noise. 
Also, since the error of \Erase{1 m/s}\Add{\SI{1}{m/s}}-\(3\sigma\) is at most \Erase{0.003 \%}\Add{\SI{0.003}{\%}} of the nominal flyby speed of \Erase{33 km/s}\Add{\SI{33}{km/s}}, the base vectors constituting STR coordinate can be determined with enough accuracy.
Therefore, the proposed method utilize STR coordinate.
Thus, the system equation of the state vector is as follows.
\begin{equation}
\label{eq:state_eq}
    \dot{\textbf{x}}  = [\textbf{v}^{T}_{(t_0)}, \textbf{0}^{T}]^{T} + \textbf{w}
\end{equation}
where, \textbf{0} represents seven\Add{-}dimensional zero vector and \(\textbf{w}\) is a \Erase{10}\Add{ten}-dimensional vector, representing the process noise taking into account the velocity error and the alignment variation.

\subsection{Observation model in navigation filter} 
The measurement value for the optical navigation is the luminance center of the target body on the image captured by the telescope. 
Assuming that the image of the target is captured at time \(t_s\), the target direction \(\textbf{d}\) with respect to the spacecraft before mirror reflection is calculated as Eq. (\ref{eq:target_dir}) in the base coordinate of the telescope.
\begin{equation}
\label{eq:target_dir}
    \textbf{d} = C_{(\phi_A,\theta_A,\phi_A)}C_{b2t}C_{i2b}\left(-\frac{\textbf{r}_{(t_s)}}{|\textbf{r}_{(t_s)}|}\right)
\end{equation}
where, 
\begin{equation}
\label{eq:rot_matrix}
    C_{(\phi_A,\theta_A,\phi_A)} =
\begin{pmatrix}
    1 & \psi_A & -\theta_A \\
    -\psi_A & 1 & \phi_A \\
    \theta_A & -\phi_A & 1
\end{pmatrix}
\end{equation}
is a coordinate transformation matrix representing the mounting error of the telescope, which corresponds to the effect of type-A misalignment described in the previous section.
\(C_{i2b}\) and \(C_{b2t}\) are also coordinate transformation matrices from the STR coordinate to the spacecraft body frame and spacecraft body frame to the designed base coordinate of the telescope, respectively.
On the onboard computer, \(C_{i2b}\) is estimated by combining the spacecraft's estimated attitude provided by AOCS and \(\textbf{v}_{(t_0)}\) determined from DDOR. 
\(\textbf{r}_{(t_s)}\) is the spacecraft's relative position at \(t = t_s\) and is obtained from Eq. (\ref{eq:lin-mortion}) on the onboard computer.
From Eq. (\ref{eq:L-missed}), the target direction after mirror reflection \(\textbf{d}^*\) is calculated as Eq. (\ref{eq:d_reflected}).
\begin{equation}
\label{eq:d_reflected}
    \textbf{d}^* = [d_{x}^*, d_{y}^*, d_{z}^*]^{T} = \textbf{d} - 2\left( \textbf{d} \cdot \textbf{n}_{(\eta)} \right) \textbf{n}_{(\eta)}
\end{equation}
As shown in Eqs. (\ref{eq:n-missed}) to (\ref{eq:L-missed}), \(\textbf{n}_{(\eta)}\) is a function of misalignments of type-B to D and the telescope rotation angle.
From the perspective projection model \cite{Golub}, the two-dimensional position of the target body projected on the imager \(\textbf{y}\) is calculated as Eq. (\ref{eq:y_pix}).
\begin{equation}
\label{eq:y_pix}
    \textbf{y} =
\begin{pmatrix}
    x_p \\
    y_p
\end{pmatrix}
    = \tfrac{f_{opt}}{d_{z}^*}
\begin{pmatrix}
    d_{x}^* \\
    d_{y}^*
\end{pmatrix}
+\boldsymbol{v}
\end{equation}
where, \(f_{opt}\) denotes the focal length of the telescope. 
\(x_p\) and \(y_p\) represent horizontal and vertical position of the centroid on the imager plane, respectively. 
\(\boldsymbol{v}\) is a two-dimensional vector, representing the observation noise.
Thus, from Eqs. (\ref{eq:lin-mortion}) -- (\ref{eq:y_pix}), the predicted measurement \(\textbf{y}\) is obtained from the estimated state vector \(\textbf{x}\) on the onboard computer.

\subsection{Filter equations} 
As described in the previous section, the mapping function from the state vector \(\textbf{x}\) to the measurement value \(\textbf{y}\) is nonlinear. 
To avoid linearization error, this study adopts the unscented Kalman filter (UKF) \cite{Crassidis}, instead of the extended Kalman filter (EKF) used in previous studies \cite{Bhaskaran, Okada, good}. 
Although EKF is advantageous in terms of computational time \cite{Gupta}, several studies reported that UKF tends to perform better \cite{Gupta}.
In addition, in the case of the $\text{DESTINY}^\text{+}$ mission, the measurement update interval is limited to \Erase{1 s}\Add{\SI{1}{s}} at the fastest due to the hardware constraint described later, so there is a relatively larger margin of time available for the observation update process.
Considering these points, this study uses UKF, instead of EKF.

In the UKF, at each measurement update timing of k-th step, a set of 2n+1 state vectors \(\chi_{k}\) called sigma points is generated from a \(n \times n\) post-updated state covariance matrix \(P_k^+\) and an n-dimensional post-updated state vector \(\hat{\textbf{x}}_k^+\), as follows
\begin{equation}
\label{eq:sigma_points}
    \chi_{k(i)}=
    \begin{cases}
        \hat{\textbf{x}}_k^+ & (i=0) \\
        \hat{\textbf{x}}_k^+ + \left[ \sqrt{(n+\lambda)P_k^+} \right]_{i} & (i=1 \sim n) \\
        \hat{\textbf{x}}_k^+ - \left[ \sqrt{(n+\lambda)P_k^+} \right]_{i-n} & (i=n+1 \sim 2n)
    \end{cases}
\end{equation}
where, “(i)” denotes i-th member of the sigma points; \(\sqrt{M}\) denotes the matrix square root, which represents a matrix A such that \(AA^t=M\). “\([M]_i\)” is intended as an operation extracting the i-th column vector from the matrix M. 
\(\lambda\) is a hyper parameter.
At each time update of the k+1-th step, all sigma points are propagated forward in the time domain \Add{in accordance with Eqs. (14) and (16), as follows. In Eq. (22), \(dt\) denotes the filter update interval}. 
\EraseL{Since this study assumes that the state vector is constant, as described in Eq. (\ref{eq:state_eq}), the time update equation of the state vector becomes} 
\begin{equation}
    \chi_{k+1(i)}=\chi_{k(i)} \Add{+ [\textbf{v}_{(t_{0})}^{T}, \textbf{0}^{T}]^{T}dt}
\end{equation}
From \(\chi_{k+1}\), the predicted mean state \(\hat{\textbf{x}}_{k+1}^-\) and predicted state covariance \(P_{k+1}^-\) are given as follows, with considering the process noise covariance matrix \(Q\).
\begin{equation}
\label{eq:mean_state}
    \hat{\textbf{x}}_{k+1}^- =\frac{1}{n+\lambda}\left( \chi_{k+1(0)} + \frac{1}{2}\sum_{i=1}^{2n} \chi_{k+1(i)}\right)
\end{equation}
\begin{equation}
\begin{split}
    P_{k+1}^- =\frac{1}{n+\lambda}
    \left\{
    \lambda 
    \left[ \chi_{k+1(0)} - \hat{\textbf{x}}_{k+1}^- \right]
    \left[ \chi_{k+1(0)} - \hat{\textbf{x}}_{k+1}^- \right]^{T} \right.\\
    \left.
    + \frac{1}{2}\sum_{i=1}^{2n}
    \left[ \chi_{k+1(i)} - \hat{\textbf{x}}_{k+1}^- \right]
    \left[ \chi_{k+1(i)} - \hat{\textbf{x}}_{k+1}^- \right]^{T}
    \right\}+Q
\end{split}
\end{equation}
Applying Eqs. (\ref{eq:lin-mortion}) -- (\ref{eq:y_pix}) to each \(\chi_{k+1}\), the predicted measurements \(\gamma_{k+1}\) are obtained. 
From the predicted measurements \(\gamma_{k+1}\), the mean predicted measurement \(\hat{\textbf{y}}_{k+1}^-\) and its covariance \(P_{k+1}^{yy}\) are obtained as follows:
\begin{equation}
    \hat{\textbf{y}}_{k+1}^- =\frac{1}{n+\lambda}\left( \gamma_{k+1(0)} + \frac{1}{2}\sum_{i=1}^{2n} \gamma_{k+1(i)}\right)
\end{equation}
\begin{equation}
\label{eq:meas_cov}
\begin{split}
    P_{k+1}^{yy} = \frac{1}{n+\lambda}
    \left\{
    \lambda 
    \left[ \gamma_{k+1(0)} - \hat{\textbf{y}}_{k+1}^- \right]
    \left[ \gamma_{k+1(0)} - \hat{\textbf{y}}_{k+1}^- \right]^{T}\right.\\
    \left.
    + \frac{1}{2}\sum_{i=1}^{2n}
    \left[ \gamma_{k+1(i)} - \hat{\textbf{y}}_{k+1}^- \right]
    \left[ \gamma_{k+1(i)} - \hat{\textbf{y}}_{k+1}^- \right]^{T}
    \right\}
\end{split}
\end{equation}
From Eqs. (\ref{eq:mean_state}) -- (\ref{eq:meas_cov}), the innovation covariance \(P_{k+1}^{vv}\) and cross-correlation matrix \(P_{k+1}^{xy}\) at the measurement update timing (k+1)-th step are obtained as follows, with considering the measurement noise covariance matrix \(R\).
\begin{equation}
\label{eq:inov_cov}
    P_{k+1}^{vv} = P_{k+1}^{yy} + R
\end{equation}
\label{eq:cross_cov}
\begin{equation}
\begin{split}
    P_{k+1}^{xy} = \frac{1}{n+\lambda}
    \left\{
    \lambda 
    \left[ \chi_{k+1(0)} - \hat{\textbf{x}}_{k+1}^- \right]
    \left[ \gamma_{k+1(0)} - \hat{\textbf{y}}_{k+1}^- \right]^{T} \right.\\
    \left.
    + \frac{1}{2}\sum_{i=1}^{2n}
    \left[ \chi_{k+1(i)} - \hat{\textbf{x}}_{k+1}^- \right]
    \left[ \gamma_{k+1(i)} - \hat{\textbf{y}}_{k+1}^- \right]^{T} 
    \right\}
\end{split}
\end{equation}
Then, the Kalman gain at (k+1)-th step \(K_{k+1}\) is computed as follows:
\begin{equation}
\label{eq:k_gain}
    K_{k+1} = P_{k+1}^{xy}\left(P_{k+1}^{vv}\right)^{-1}
\end{equation}
\begin{figure*}[!tpb]
	\centering
	\includegraphics[scale=0.65]{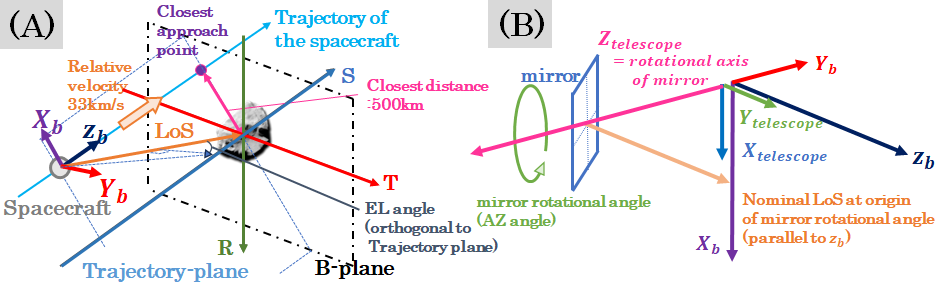}
	\caption{Geometrical conditions in simulation scenario
                (A) flyby geometry (relationship between spacecraft body coordinate and STR coordinate),
                (B) alignment of rotating telescope relative to spacecraft body coordinate}
	\label{fig:geometry_in_SCLT}
\end{figure*}
\begin{table*}[!t]
\centering
\caption{Parameters in simulation scenario}\label{tbl_sim_config}
\begin{tabular}{l p{7cm} l p{4.5cm} }
\hline
ID & \Erase{p}\Add{P}arameter name & \Erase{v}\Add{V}alue & \Erase{n}\Add{N}ote\\
\hline
1& Nominal relative distance on B-plane & \Erase{500 km}\Add{\SI{500}{km}} &  \\
2& Nominal relative velocity on S-axis & \Erase{33 km/s}\Add{\SI{33}{km/s}} &  \\
3& Initial relative position error in STR & \Erase{90 km}\Add{\SI{90}{km}}-3\(\sigma\) & in T,R axis \\
 & & \Erase{130 km}\Add{\SI{130}{km}}-3\(\sigma\) & in S-axis \\
4& Relative velocity error in STR &  \Erase{1 m/s}\Add{\SI{1}{m/s}}-3\(\sigma\) \\
5& Observation error of asteroid position on imager & \Erase{0.006 deg}\Add{\SI{0.006}{deg}}-3\(\sigma\) & same for vertical and horizontal \\
6& Misalignment & \Erase{4.6e-3 deg}\Add{\SI{4.6e-3}{deg}}-3\(\sigma\) & for ${\phi_A}$,${\theta_A}$,${\psi_A}$ \\
 & & \Erase{0.01 deg}\Add{\SI{0.01}{deg}}-3\(\sigma\) & for ${\phi_B}$,${\delta_B}$ \\
 & & \Erase{0.01 deg}\Add{\SI{0.01}{deg}}-3\(\sigma\) & for ${\delta_C}$,${\delta_D}$ \\ 
\hline
\end{tabular}
\end{table*}
From Eqs. (\ref{eq:inov_cov}) -- (\ref{eq:k_gain}), the measurement update equation of the state vector and its covariance matrix is obtained as follows:
\begin{eqnarray}
    \hat{\textbf{x}}_{k+1}^+ &=& \hat{\textbf{x}}_{k+1}^- + K_{k+1}\left(\textbf{y}_{k+1} - \hat{\textbf{y}}_{k+1}^-\right) \\
    P_{k+1}^{+} &=& P_{k+1}^{-} - K_{k+1}P_{k+1}^{vv}K_{k+1}^t
\end{eqnarray}
where, \(\textbf{y}_{k+1}\) denotes the actual measurement value obtained from the image at the (k+1)-th step. After the measurement update, a new set of sigma points is generated according to Eq. (\ref{eq:sigma_points}) by replacing $k \rightarrow k+1$.

\section{Numerical Simulation on PC} 

To evaluate the effectiveness of the proposed algorithm, a numerical simulation was performed on a PC. 
The simulation was conducted using the Monte-Carlo method with multiple runs under varying error conditions. 
This section describes the simulation scenario and conditions, followed by a discussion on the effectiveness and limitations of the proposed algorithm based on the simulation results.

\subsection{Simulation scenario and conditions on PC} 
Based on the timing for starting the onboard autonomous navigation in the $\text{DESTINY}^\text{+}$ mission, the simulation scenario starts \Erase{3600 s}\Add{\SI{3600}{s}} before the spacecraft's closest approach to the target body.
At this point, it is assumed that the initial estimation values for the spacecraft's trajectory relative to the target have been obtained through a combination of DDOR and ground-in-the-loop optical navigation. 
At the start of the simulation, the onboard optical navigation begins to estimate and correct the errors in the initial estimation values of the spacecraft's relative trajectory. This is achieved by utilizing the centroid position of the target in the images captured by the rotating telescope at 1-s intervals throughout the simulation.
This interval derives from the hardware limitation of the image processor on TCAP. 
After the image is captured, the processor reduces the noise in the image and detects the centroid position of the asteroid on the imager.
Since the process requires about \Erase{1 s}\Add{\SI{1}{s}} to be executed, the shutter interval is \Erase{1 s}\Add{\SI{1}{s}} at fastest.
The feasibility of this interval in terms of the time margin for the navigation computation time is evaluated in section 5 with utilizing real hardware.
\begin{table*}[!t]
\centering
\caption{Filter parameters}\label{tbl_filter_param}
\begin{tabular}{l p{7cm} l p{7cm} }
\hline
\Erase{p}\Add{P}arameter name & \Erase{v}\Add{V}alue & \\
\hline
Initial state covariance \(P_{0}\) & \(diag(45^{2}, 30^{2}, 30^{2}\) & [\(\rm Km^2\)] \\
& ~~~~~~~\(1.5^{2}, 1.5^{2}, 1.5^{2}, 3.4^{2}, 3.4^{2}, 3.4^{2}, 3.4^{2})\) & [\(\rm mdeg^2\)] \\
Process noise covariance \(Q\) & \(diag\)(1e-6, 1e-6, 1e-6, & [\(\rm km^2\)] \\
& ~~~~~~~1e-6, 1e-6, 1e-6, 1e-6, 1e-6, 1e-6, 1e-6) & [\(\rm mdeg^2\)] \\
Observation noise covariance \(R\) & \(diag(8.0^2, 8.0^2)\) & [\(\rm mdeg^2\)] \\
\Add{UKF hyper parameter \(\lambda\)} & \Add{-7} & \Add{[ - ]}\\
\hline
\end{tabular}
\end{table*}
\begin{table*}[!t]
\centering
\caption{Evaluated navigation algorithms}\label{tbl_filters}
\begin{tabular}{l p{8cm} l p{7cm} }
\hline
\Erase{f}\Add{F}ilter name & \Erase{d}\Add{D}escription & \Erase{s}\Add{S}tate vector\\
\hline
PositionOnly & estimate only relative position & \(\boldsymbol{x} = \boldsymbol{r}\) \\
BiasModel & estimate relative position and two bias error angle of LoS & \(\boldsymbol{x} = [\boldsymbol{r}^{T}, \phi_{A}, \psi_{A}]^{T}\)\\
Proposed     & estimate relative position and detailed misalignments & Eq. (\ref{eq:state_vec})\\
\hline
\end{tabular}
\end{table*}

In the scenario, the target body is tracked by controlling the LoS direction with cooperative control of the spacecraft’s attitude and TCAP rotational angle, utilizing the estimation results on the relative motion of the target body provided by autonomous optical navigation.
Figs. \ref{fig:geometry_in_SCLT}-(A) and \ref{fig:geometry_in_SCLT}-(B) describe the flyby geometry and alignment of the rotating telescope relative to the spacecraft’s body in the simulation scenario, respectively.
In cooperative control, the AOCS controls the LoS direction in the elevational (EL) direction, while the TCAP controller controls the LoS direction in the azimuthal (AZ) direction. 
In the $\text{DESTINY}^\text{+}$ mission, the rotational axis of TCAP should be aligned to the perpendicular direction of the trajectory plane with the spacecraft's attitude maneuver based on the spacecraft's estimated position on the T- and R-axes obtained from the optical navigation, at least five minutes before the closest approach. 
Therefore, in the following, the accuracy of the optical navigation at five minutes before the closest approach is evaluated.

The detailed parameters in the simulation scenario are summarized in Table \ref{tbl_sim_config}. 
In this table, the nominal trajectory and the initial errors on the relative position and velocity are set based on the Phaethon flyby case in the $\text{DESTINY}^\text{+}$ mission.
\Add{The observation error is set based on the ground measurement results of the TCAP imaging system's dark noise and readout noise. 
The type-B and type-C misalignment values are set based on the accuracy of the ground calibration equipment for TCAP and mainly derive from machining accuracy of the equipment. 
On the other hand, the type-A misalignment values are set based on the accuracy of the on-ground spacecraft's system alignment calibration and mainly derive from the star tracker's accuracy used in the DESTINY+ mission.}
In the simulation, the relative trajectory of the spacecraft with respect to the target is modeled as linear motion in STR coordinates with constant velocity.
The spacecraft's initial position, velocity, and observation errors were randomly changed in the Monte Carlo simulation according to the standard deviations shown in Table \ref{tbl_sim_config}.

Also, filter parameters used in this analysis are summarized in Table \ref{tbl_filter_param}.
In the initial state covariance \({P}_{0}\), the first three components correspond to the initial position error.
The next three components correspond to type-A misalignments, and the last four components correspond to type-B to D misalignments.
\Add{Throughout the simulation, the covariance matrices Q and R are constant at the values listed in the table.}
These values are determined from the simulation settings described in Table \ref{tbl_sim_config}.

In the following, the performance of three different types of optical navigation algorithms are compared and evaluated.
The outline of each algorithm is almost the same, but the state vector is modified to compare the robustness against the misalignments.
Table \ref{tbl_filters} summarizes the navigation algorithms evaluated in the following.
The “PositionOnly" filter estimates only relative position of the spacecraft.
In this filter, the state vector is only the first three components of Eq. (\ref{eq:state_vec}) and the initial state error covariance is \(3 \times 3\) diagonal matrix consists of the first three diagonal terms of \(P_{0}\) in Table \ref{tbl_filter_param}. The “BiasModel" filter estimates relative position and two bias error angles of the LoS similar to the previous study \cite{good}, but utilizes UKF instead of EKF.
In this filter, the state vector is relative position and \(\phi_{A}\), \(\psi_{A}\) of Eq. (\ref{eq:state_vec}) and the initial state error covariance is \(5 \times 5\) diagonal matrix consists of first five diagonal terms of \(P_{0}\) in Table \ref{tbl_filter_param}.
The “Proposed" filter is the proposed method described in the previous section. 

\subsection{Simulation results on PC} 

\subsubsection{Misalignment-induced accuracy degradation} 

To assess the degree of degradation of the optical navigation accuracy due to the misalignment, the performance of “PositonOnly" filter is evaluated under two different simulation conditions.
In the first condition, all misalignments described in Table \ref{tbl_sim_config} (ID 6) are set to zero to eliminate any misalignment effect (case 0 condition). 
Then, a simulation with all misalignments added, represented as ID 6 in Table \ref{tbl_sim_config} (case 1 condition), was performed.

\begin{figure}[!t]
    \centering
    \includegraphics[scale=0.83]{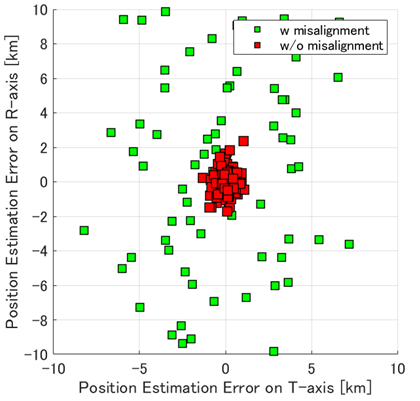}
    \caption{Distribution of position estimation errors at five minutes before closest approach obtained through 100 trials with “PositionOnly" filter}
    \label{fig:EKF_ERR}
\end{figure}

Fig. \ref{fig:EKF_ERR} describes the distributions of the estimation error on the spacecraft position projected on the B-plane obtained thorough 100 trials with “PositionOnly" filter.
The error is evaluated at five minutes before the closest approach.
In this figure, the red squares represent the distribution of the estimation errors in the case 0 condition, while the green squares represent those in case 1. 
As can be seen from this figure, the distribution of the green squares is more extensive than that of the red squares. 
The standard deviation (1-\(\sigma\)) of the distribution of the red squares is \Erase{0.4 km}\Add{\SI{0.4}{km}}, while that of the green squares is \Erase{4.6 km}\Add{\SI{4.6}{km}}. 
These results suggest that the optical navigation accuracy is degraded due to the misalignment of the rotational telescope without the application of a mitigation technique.

\subsubsection{Mitigating accuracy degradation} 

\begin{figure}[!t]
    \centering
    \includegraphics[scale=.82]{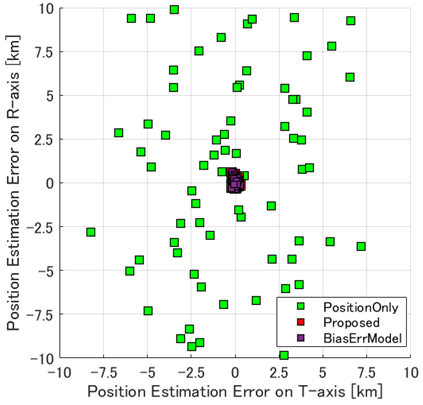}
    \caption{Comparison of the distribution of the spacecraft position estimation errors with three different filters at five minutes before closest approach obtained through 100 trials}
    \label{fig:UKF_ERR}
\end{figure}

The performance of “BiasModel” filter and “Proposed" filter were investigated under the case 1 condition. 
Fig. \ref{fig:UKF_ERR} compares the distributions of the spacecraft position estimation errors by the three navigation filters, plotted in the same format as Fig. \ref{fig:EKF_ERR}.
In this figure, the red squares represent the distribution of the estimation error with “Proposed" filter, the purple squares represent the error distribution of “BiasModel" filter, while the green squares represent the error distribution of “PositionOnly" filter as in Fig. \ref{fig:EKF_ERR}. 
As can be seen from this figure, the distribution of the red squares and the purple squares are more intensive than that of the green squares.
The standard deviation (1-\(\sigma\)) of the distribution of the red squares and the purple squares are both \Erase{0.11 km}\Add{\SI{0.11}{km}}, while that of the green squares is \Erase{4.62 km}\Add{\SI{4.62}{km}}. 
The standard deviation of the distribution of the red squares and the purple squares in this figure is almost the same as that of the \Erase{yellow}\Add{red} squares in Fig. \ref{fig:EKF_ERR}.
The accuracy of each filter at five minutes before the closest approach obtained thorough 100 trials is summarized in Table \ref{tbl_filter_compare}.
This result suggests that, under the presence of the misalignment, both “Proposed" and “BiasModel" filters can mitigate the misalignment effect to an accuracy comparable to that of “PositionOnly" filter in the absence of misalignment. 
Hence, both “Proposed" and “BiasModel" filters successfully mitigate the misalignment-induced degradation of navigation accuracy, under the assumed misalignment condition in the $\text{DESTINY}^\text{+}$ mission.

\begin{table}[!t]
\caption{\Erase{Summery} \Add{Summary} of three filters' navigation accuracy projected on B-plane}\label{tbl_filter_compare}
\centering
\begin{tabular}{p{2cm} p{1.7cm} p{2.7cm}}
\hline
Filter name & Accuracy & Note \\
\hline
PositionOnly & \Erase{0.40 km}\Add{\SI{0.40}{km}}-1\(\sigma\) & w/o misalignment\\
PositionOnly & \Erase{4.62 km}\Add{\SI{4.62}{km}}-1\(\sigma\) & w/ misalignment\\
BiasModel    & \Erase{0.11 km}\Add{\SI{0.11}{km}}-1\(\sigma\) & w/ misalignment\\
Proposed     & \Erase{0.11 km}\Add{\SI{0.11}{km}}-1\(\sigma\) & w/ misalignment\\
\hline
\end{tabular}
\end{table}

\begin{figure*}[!t]
    \centering
    \includegraphics[scale=0.6]{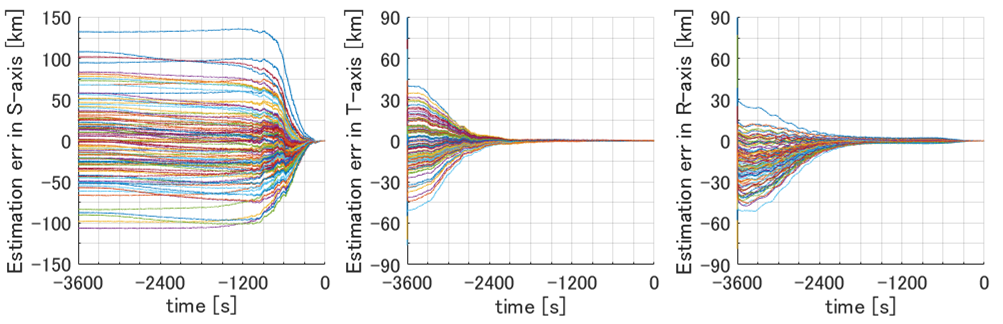}
    \caption{History of position estimation errors with “Proposed” filter obtained through 100 trials}
    \label{fig:10dimUKF_MC_History}
\end{figure*}
\begin{figure*}[!t]
    \centering
    \includegraphics[scale=0.6]{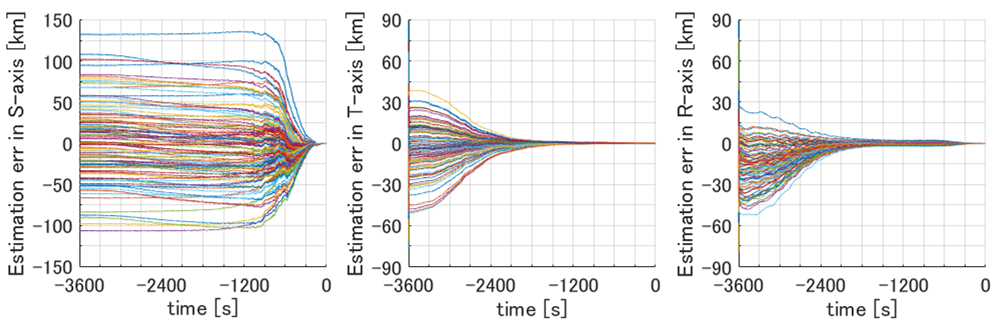}
    \caption{History of position estimation errors with “BiasModel” filter obtained through 100 trials}
    \label{fig:5dimUKF_MC_History}
\end{figure*}
\begin{figure}[!t]
    \centering
    \includegraphics[scale=0.6]{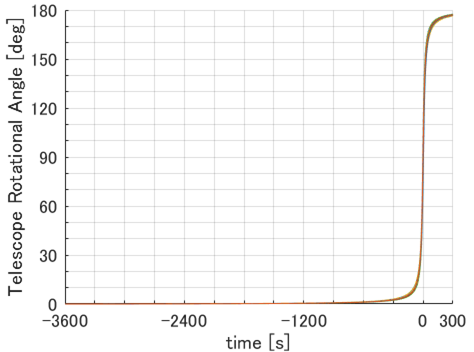}
    \caption{History of telescope rotational angle obtained through 100 trials}
    \label{fig:telescope_angle}
\end{figure}

\subsubsection{Comparison of “Proposed" and “BiasModel"}  

\begin{figure*}[!t]
    \centering
    \includegraphics[scale=0.6]{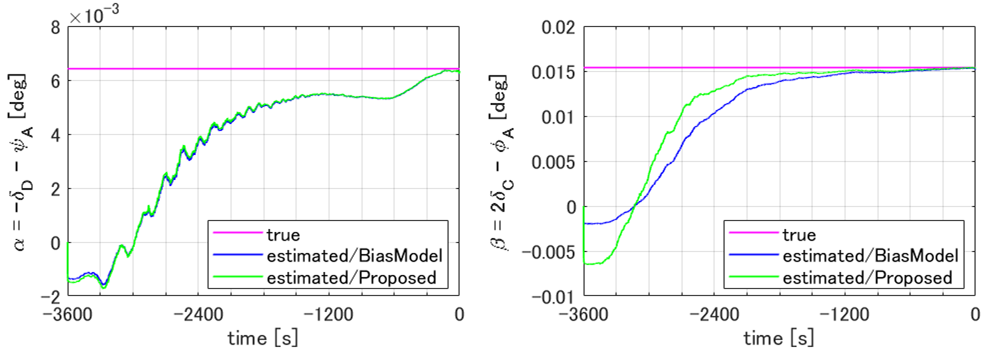}
    \caption{Example history of LoS error angle (\(\alpha\), \(\beta\)) estimation taken from 100 trials}
    \label{fig:misalignment_est_history}
\end{figure*}
\begin{figure*}[!t]
    \centering
    \includegraphics[scale=0.6]{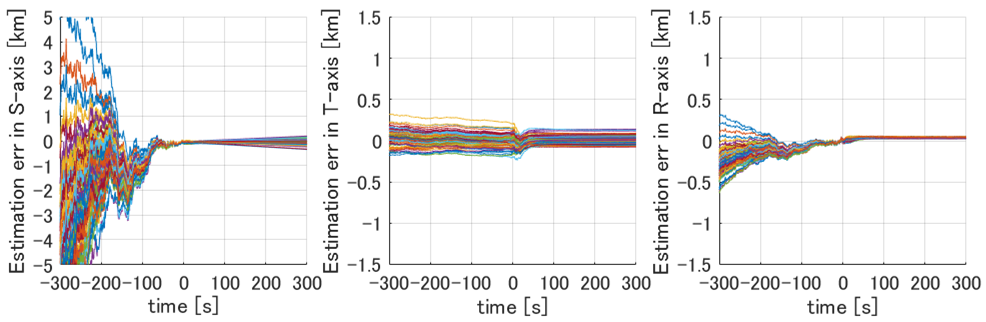}
    \caption{History of position estimation errors with “Proposed” filter obtained through 100 trials focusing on 300 s before and after the closest approach}
    \label{fig:10dimUKF_MC_History_zoomup}
\end{figure*}
\begin{figure*}[!t]
    \centering
    \includegraphics[scale=0.6]{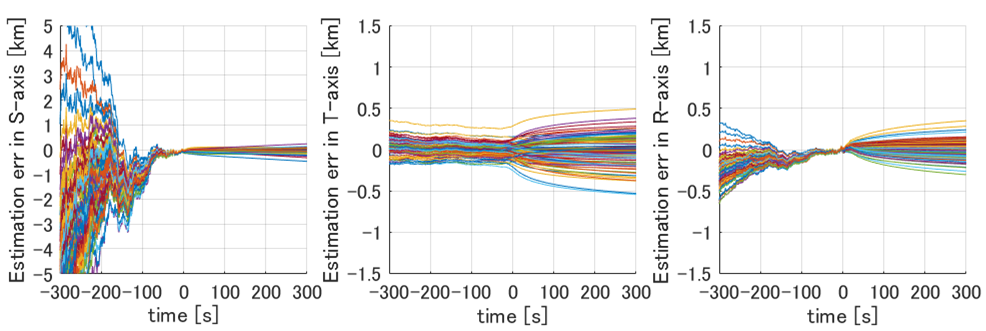}
    \caption{History of position estimation errors with “BiasModel” filter obtained through 100 trials focusing on 300 s before and after the closest approach}
    \label{fig:5dimUKF_MC_History_zoomup}
\end{figure*}

For a more detailed comparison of “Proposed” filter and “BiasModel” filter, the position estimation error histories of the two filters over the 100 trials are described in Figs. \ref{fig:10dimUKF_MC_History} and \ref{fig:5dimUKF_MC_History}, respectively. In these figures, the horizontal axes represents the time from the closest approach and the time evolution of the estimation error for each of the 100 trials is plotted in different colors.
Comparing these two figures, there is little difference between them, although the estimation error in T-axis direction decreases slightly faster in "Proposed" filter.
This result suggests that, under the assumed simulation condition, there are little difference in performance of the navigation accuracy between “Proposed” and “BiasModel”, even though “Proposed” filter uses more detailed misalignment model.
This result derives from the time history of the telescope rotational angle. Fig. \ref{fig:telescope_angle} describes the telescope rotational angle during the 100 trials, plotted in the same format as Fig. 15.
As described in the figure, the rotational angle of the rotating telescope is approximately zero until a few minutes before the closest approach timing, since the relative direction of the target asteroid does not change significantly until that timing. 
It means that \(\eta\) in Eq. (\ref{eq:L-missed}) approximately equals \(\delta_D\) for most of the simulation duration.
In this case, considering the geometrical condition (Fig. \ref{fig:geometry_in_SCLT}) and Eqs. (\ref{eq:target_dir}), (\ref{eq:rot_matrix}), the LoS direction of the telescope in the spacecraft's body frame becomes as follows.
\begin{equation}
\label{eq:L-missed_zero}
\begin{split}
    &\textbf{L}_{b} =  {C^{T}_{b2t}}{C^{T}_{(\phi_A,\theta_A,\phi_A)}}\textbf{L}\\
    &\approx
\begin{pmatrix}
    1 & 0 & 0 \\
    0 & 0 & -1 \\
    0 & 1 & 0
\end{pmatrix}
\begin{pmatrix}
    1 & -\psi_A & \theta_A \\
    \psi_A & 1 & -\phi_A \\
    -\theta_A & \phi_A & 1
\end{pmatrix}
\begin{pmatrix}
    -\delta_D \\
    1\\
    -2\delta_C
\end{pmatrix}
\\
&\approx
[-\delta_D - \psi_A, 2\delta_C - \phi_A, 1]^{T}
\end{split}
\end{equation}
Eq. (\ref{eq:L-missed_zero}) suggests that the misalignment components \(\theta_A\), \(\delta_B\), and \(\phi_B\) do not contribute to the LoS direction of the telescope when the rotational angle of the telescope is approximately zero.
Thus, the LoS error can be described with two bias angles \(\alpha(=-\delta_{D}-\psi_{A})\) and \(\beta(=2\delta_{C}-\phi_{A})\), which can be handled with the “BiasModel" filter.
Fig. \ref{fig:misalignment_est_history} describes an example of the estimation history of \(\alpha\) and \(\beta\) extracted from the 100 trials.
In this figure, the magenta line represents the true value of \(\alpha\) and \(\beta\), the blue and green lines represent the estimated value with “BiasModel" filter and “Proposed" filter, respectively.
In “BiasModel", \(\alpha\) is obtained as the estimation result of \(-\psi_{A}\) and \(\beta\) as the estimation result of \(-\phi_{A}\).
In “Proposed", \(\alpha\) is computed from the estimation result of \(\delta_{D}\) and \(\psi_{A}\), while \(\beta\) is computed from the estimation result of \(\delta_{C}\) and \(\phi_{A}\).
As described in this figure, both of “BiasModel" filter and “Proposed" filter can estimate \(\alpha\) and \(\beta\) correctly.
Therefore, there is no difference in navigation accuracy between the two filters until just before the closest approach, when the telescope rotation angle is small.
\begin{figure*}[!t]
    \centering
    \includegraphics[scale=0.65]{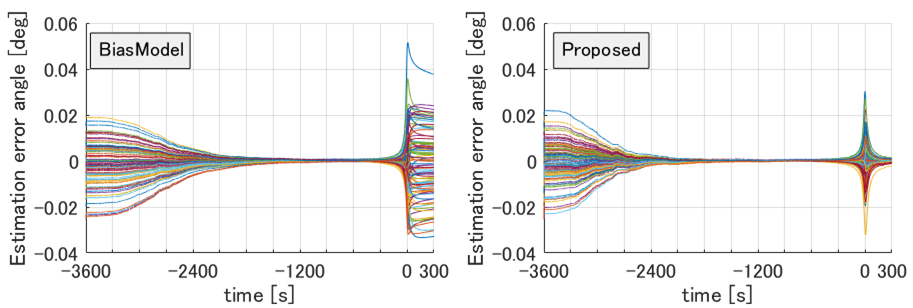}
    \caption{History of LoS estimation error angle obtained through 100 trials}
    \label{fig:LoS_Est_Error_Compare}
\end{figure*}

However, during and immediately after the closest approach, when the telescope rotation angle becomes large, there is a difference in navigation performance between the two filters.
Figs. \ref{fig:10dimUKF_MC_History_zoomup} and \ref{fig:5dimUKF_MC_History_zoomup} describe again the position estimation error histories of the two filters over the 100 trials, as in Figs. \ref{fig:10dimUKF_MC_History} and \ref{fig:5dimUKF_MC_History}, but focusing on \Erase{300 s}\Add{\SI{300}{s}} before and after the closest approach timing.
Comparing these figures, it can be seen that the navigation error increases around the closest approach time for “BiasModel" filter, whereas the navigation accuracy is maintained for “Proposed" filter.
This difference is attributed to the behavior of the LoS error. As described in Fig. \ref{fig:telescope_angle}, the telescope rotation angle rapidly changes around the closest approach time. 
In this case, as described in section 2.3, the LoS error changes depending on the rotational angle, which cannot described as constant bias angles. 
Therefore, it is difficult for “BiasModel" filter to track changes in the LoS error, resulting in a degradation of navigation accuracy.
By contrast, “Proposed" filter estimates the misalignment parameters for describing the LoS error, which makes it easier to follow changes in the LoS error and thus prevents degradation of navigation accuracy.
To confirm this point, Fig. \ref{fig:LoS_Est_Error_Compare} plots the histories of the LoS direction estimation error with both filter for 100 trials.
For both filters, the estimation error decreases to nearly zero from the start of the simulation to just before the closest approach timing, but the error increases rapidly just before the closest approach timing when the telescope rotation angle changes rapidly.
After the closest approach timing, the estimation error remains large in “BiasModel" filter's history, while it becomes small again for “Proposed" filter's history, and this difference also affects the difference in navigation accuracy appears in Figs. \ref{fig:10dimUKF_MC_History_zoomup} and \ref{fig:5dimUKF_MC_History_zoomup}.
This result suggests that the estimation error before and after a large rotation of the telescope can be mitigated by estimating each alignment error parameter that causes the LoS error, instead of estimating the LoS error directly as an instantaneous bias angle.

\begin{table}[!t]
\caption{Comparison of “BiasModel" and “Proposed" at 300s after closest approach}\label{tbl_filter_compare2}
\begin{tabular}{p{2cm} p{2cm} p{2cm}}
\hline
Filter name & Accuracy & Note \\
\hline
BiasModel & \Erase{0.115 km}\Add{\SI{0.115}{km}}-1\(\sigma\)  & position error\\
Proposed  & \Erase{0.021 km}\Add{\SI{0.021}{km}}-1\(\sigma\)  & on B-plane\\
BiasModel & \Erase{0.014 deg}\Add{\SI{0.014}{deg}}-1\(\sigma\) & LoS error\\
Proposed  & \Erase{0.005 deg}\Add{\SI{0.005}{deg}}-1\(\sigma\) & LoS error\\
\hline
\end{tabular}
\end{table}

Table. \ref{tbl_filter_compare2} compares the position and the LoS estimation error of both “BiasModel" and “Proposed" filters at \Erase{300 s}\Add{\SI{300}{s}} after closest approach.
Although there is little difference between the two filters in terms of the accuracy of position estimation just before the closest approach, as shown in this table, “Proposed" filter has a higher estimation accuracy immediately after the closest approach.
This result suggests that incorporating a detailed misalignment model into the navigation filter will enable more robust navigation against the misalignments.

\section{Hardware-In-the-Loop Simulation (HILS)} 

To validate the applicability of the proposed algorithm for real missions, the algorithm was implemented as flight software on the breadboard model of the OBC for the $\text{DESTINY}^\text{+}$ mission. 
The OBC was connected to a dynamics simulator running on a real-time computer to perform HILS. 
In the HILS, the performance of the algorithm running on the OBC was evaluated considering the following two points:
(1) navigation accuracy compared to that of the algorithm running on PC described in the previous section, 
(2) computational cost (required computation time). 
This section describes the overview of the system configuration of the HILS and the evaluation results of the algorithm as flight software.

\subsection{Simulation condition of HILS} 

\begin{figure}[!b]
    \centering
    \includegraphics[scale=.7]{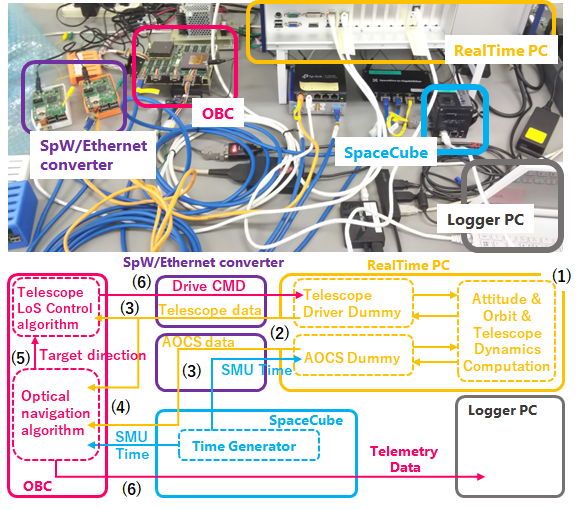}
    \caption{Overview of HILS system configuration}
    \label{fig:HILS_Config}
\end{figure}

Fig. \ref{fig:HILS_Config} illustrates the system configuration of the HILS. 
The simulation system comprised the OBC, real-time computer, logger PC, SpaceCube (SpC) \cite{Hihara} and Space Wire (SpW) \cite{Parkes} / Ethernet converters.
Although the OBC is breadboard model, its main CPU and memory size are the same as the flight model under development for the $\text{DESTINY}^\text{+}$ mission.
The real-time computer simulates the behavior of the attitude and orbit control onboard computer (AOBC) and spacecraft's attitude and orbit dynamics. Also, the computer simulates the behavior of the driver electronics and motor dynamics of TCAP.
To emulate the data flow in the real spacecraft system, including the data protocol and the data send/recieve timing, the real-time computer is connected to the OBC via the SpW/Ethernet converter and the SpC.
The details of the data flow in the HILS system is as follows.

\begin{enumerate}
    \item[1)] 
    The real-time computer calculates the dynamics of the spacecraft, such as orbit and attitude, as well as the mechanical motion of the rotating telescope.

    \item[2)] 
    Based on the calculation results, the real-time computer transmits the source data to create telemetry data to a SpW/Ethernet converter in \Erase{1 Hz}\Add{\SI{1}{Hz}} cycles. 
    The source data consists of the bright center position of the asteroid captured by the imager, the rotation angle of the rotating telescope, and the attitude data of the spacecraft.

    \item[3)] 
    The transmitted data from the real-time computer is converted into telemetry data in accordance with the SpW packet format on the SpW/Ethernet converter and then passed to the SpC. 
    The SpC controls the data flow between the spacecraft’s attitude/orbit control system (AOCS) and the OBC, as well as generates/transmits the timing signals, called spacecraft management unit time (SMU time), to synchronize onboard timers of all components.

    \item[4)] 
    After receiving the centroid position and attitude data from the SpW/ Ethernet converter, the SpC transmits these data to the OBC on a SpW line in \Erase{1 Hz}\Add{\SI{1}{Hz}} cycle.

    \item[5)] 
    After receiving these data, the OBC estimates the relative position of the spacecraft to the asteroid. 
    The OBC also calculates the drive command to control and align the LoS direction of the rotating telescope toward the relative direction of the target asteroid.

    \item[6)] 
    The calculated drive command is transmitted to the real-time computer through another SpW/Ethernet converter. 
    The log data on the calculation process on the OBC is also transmitted from the OBC to the logger PC through the SpC in \Erase{1 Hz}\Add{\SI{1}{Hz}} cycle.
\end{enumerate}
The OBC runs with a system clock of \Erase{60 MHz}\Add{\SI{60}{MHz}} and has 128 Mbyte RAM as its working memory. 
All the onboard applications including the optical navigation are driven in the multi-thread process with \Erase{1 Hz}\Add{\SI{1}{Hz}} cycle from T-Kernel, which is a type of real-time operating system running on the OBC.

The simulation scenario used in the HILS is the same as that in Table \ref{tbl_sim_config} described in the previous section.

\subsection{Simulation results on HILS} 

\begin{figure}[!t]
    \centering
    \includegraphics[scale=0.7]{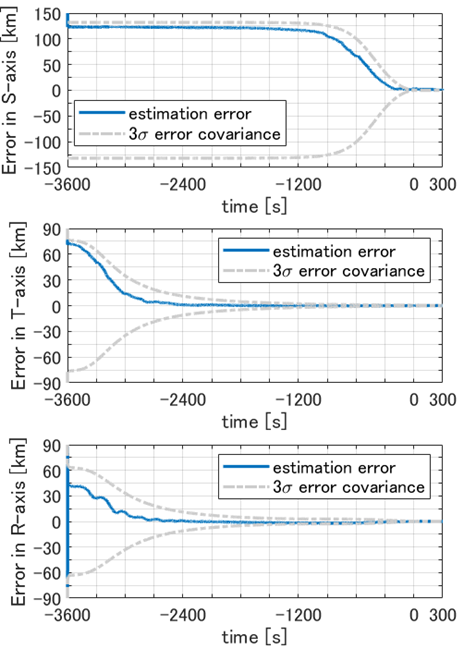}
    \caption{Example history of position estimation errors and estimation error covariances in HILS}
    \label{fig:HILS_pos_err_history}
\end{figure}

Fig. \ref{fig:HILS_pos_err_history} describes the example history of the optical navigation accuracy during the simulation scenario in the HILS. 
Comparing this figure with Fig. \ref{fig:10dimUKF_MC_History} described in the previous section, the estimation errors and error covariances show the same trend in both figures. 
The estimation error on the relative position at five minutes before the closest approach timing in the HILS is \Erase{0.2 km}\Add{\SI{0.2}{km}} on the B-plane, which is in range of the error distribution of the Monte-Carlo simulation on the PC described in section 4.2.2 (\Erase{0.33 km}\Add{\SI{0.33}{km}}-3\(\sigma\)).
These results suggest that the algorithm runs on the OBC properly.
\begin{figure}[!t]
    \centering
    \includegraphics[scale=.6]{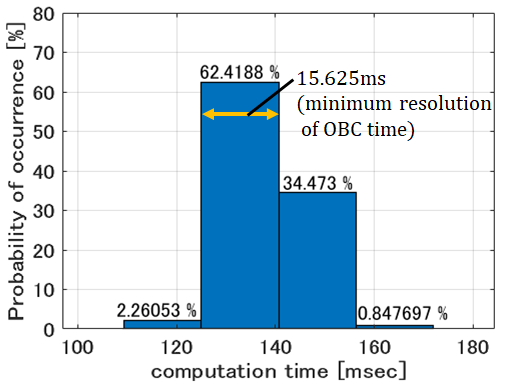}
    \caption{Histogram of required computation time for the proposed 
optical navigation algorithms during simulation. (The binwidth is limited by the minimum time resolution of OBC time, 15.625 ms.)}
    \label{fig:HILS_time_hist}
\end{figure}

Fig. \ref{fig:HILS_time_hist} describes the histogram of the computation time required to execute one measurement update process of the optical navigation algorithm on the OBC during the simulation scenario. 
The required computation time is calculated on the OBC by taking the difference between the calculation start time and the calculation end time with a time resolution of \Erase{15.625 ms}\Add{\SI{15.625}{ms}} (\Erase{64 Hz}\Add{\SI{64}{Hz}}). 
This calculation is performed every time the optical navigation algorithm runs.
Since \Erase{15.625 ms}\Add{\SI{15.625}{ms}} (\Erase{64 Hz}\Add{\SI{64}{Hz}}) is the minimum resolution of the OBC time, the minimum binwidth in Fig. \ref{fig:HILS_time_hist} is limited to \Erase{15.625 ms}\Add{\SI{15.625}{ms}} (\Erase{64 Hz}\Add{\SI{64}{Hz}}).
As indicated in Fig. \ref{fig:HILS_time_hist}, the calculation process of the optical navigation completes within approximately \Erase{0.125 s}\Add{\SI{0.125}{s}} to \Erase{0.155 s}\Add{\SI{0.155}{s}}, with a probability of higher than \Erase{90 \%}\Add{\SI{90}{\%}}. 
Even in the worst case, the calculation process requires less than \Erase{0.18 s}\Add{\SI{0.18}{s}}, which is significantly shorter than the execution cycle of each thread on the OBC (\Erase{1 Hz}\Add{\SI{1}{Hz}}). 
Therefore, Fig. \ref{fig:HILS_time_hist} indicates that the computational burden of the proposed algorithm is acceptable for real missions, since the algorithm successfully runs on the real OBC for flight.

\section{Conclusion} 

This paper proposes an autonomous optical navigation algorithm that can mitigate the accuracy degradation due to the misalignment of the rotating telescope. 
Although the misalignment was calibrated before starting the autonomous optical navigation in past flyby missions, in the case of small spacecraft missions, such as {$\text{DESTINY}^\text{+}$} mission, calibrating the navigation camera with sufficient accuracy before the closest approach is not guaranteed to be achievable due to the limited link capacity of the spacecraft. 
Therefore, in the proposed algorithm, the misalignment of the telescope is estimated simultaneously with the spacecraft’s orbit relative to the flyby target during the closest approach. 
A similar method, in which the LoS error is estimated simultaneously with the relative position as the bias angle, has been proposed in a previous study.
However, it was found that the LoS error and navigation accuracy is degraded during and after the closest approach, when the telescope rotational angle changes rapidly.
Therefore, in the proposed method, to express the LoS error as a function of the telescope rotation angle, seven parameters representing telescope misalignment were employed, and these parameters were estimated simultaneously with the relative position.
Such approach can reduce the time variation of the estimation parameters and is robust to changes in the LoS error caused by fast rotational motion of the telescope at closest approach.

The effectiveness of the proposed algorithm was evaluated via a numerical simulation on a PC, taking the Phaethon flyby in the {$\text{DESTINY}^\text{+}$} mission as an example.
In the example case, the misalignment-induced navigation accuracy degradation of \Erase{4.6 km}\Add{\SI{4.6}{km}}-\(1\sigma\) before the closest approach timing can be reduced to \Erase{0.1 km}\Add{\SI{0.1}{km}}-\(1\sigma\) by the proposed method. In addition, simulation results show that the proposed method can reduce the LoS estimation error during and after the closest approach to about half that of the method that estimates the LoS error as a bias, thus reducing the degradation of navigation accuracy after closest approach.
To verify the practicality of the purposed method as a flight software, the computational burden of the proposed algorithm running on the OBC was evaluated via HILS. 
The required time to run one-cycle of the navigation process on the onboard computer for the {$\text{DESTINY}^\text{+}$} mission is less than \Erase{0.18 s}\Add{\SI{0.18}{s}}, which is shorter than the execution cycle of the thread on the OBC.

Thus, it can be concluded that the proposed method can mitigate the misalignment-induced accuracy degradation of the optical navigation within a reasonable computational cost that can be handled within the onboard computers’ capacity.

\section*{Acknowledgment}

The authors express their gratitude to MEISEI ELECTRIC CO., LTD. and Genesia Corporation for their contribution to the development of TCAP.

\appendix
\section{Definition of STR Coordinate}
\label{app1}
\begin{figure}[!t]
    \centering
    \includegraphics[scale=.7]{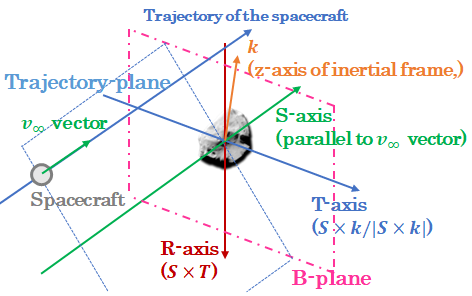}
    \caption{Definition of STR coordinate}
    \label{fig:STR_Definition}
\end{figure}

The STR coordinate is a right-handed coordinate system, which is related to the B-plane.
The origin of the coordinate is the center of the target body and consists of three orthogonal unit vectors \(\textbf{S}\), \(\textbf{T}\) and \(\textbf{R}\) defined as follows:
\begin{equation}
\label{eq:def_S}
    \textbf{S} = \textbf{v}_{\infty}/|\textbf{v}_{\infty}|
\end{equation}
\begin{equation}
\label{eq:def_T}
    \textbf{T} = \textbf{S} \times \boldsymbol{k} / |\textbf{S} \times \boldsymbol{k}|
\end{equation}
\begin{equation}
\label{eq:def_R}
    \textbf{R} = \textbf{S} \times \textbf{T}
\end{equation}
where, \(\textbf{v}_{\infty}\) is the hyperbolic excess speed vector of the spacecraft approaching to the target body, \(\boldsymbol{k}\) is an unit vector parallel to the z-axis of the inertial reference frame (typically, normal direction to the ecliptic plane \cite{Jah2002, Vallado}). Fig. \ref{fig:STR_Definition} illustrates the definition of the coordinate.

\bibliographystyle{elsarticle-num} 
\bibliography{cas-refs}






\end{document}